\documentclass[12pt,a4paper]{article}
\usepackage{setspace}
\usepackage{sectsty}

\usepackage{graphics}

\sectionfont{\normalsize\centering}

\newcommand{\addr}{\it\normalsize}

\begin{document}
\title{$ab$-plane tunneling and Andreev spectroscopy of superconducting gap and pseudogap in
(Bi,Pb)$_2$Sr$_2$Ca$_2$Cu$_3$O$_{10+\delta}$ and Bi$_2$Sr$_2$CaCu$_2$O$_{8+\delta}$}

\author{A. I. D'yachenko$\mathrm{^a}$, \and V. Yu. Tarenkov$\mathrm{^a}$, \and R. Szymczak$\mathrm{^b}$,
\and H. Szymczak$\mathrm{^b}$, \and A. V. Abal'oshev$\mathrm{^{b,}}$\thanks{abala@ifpan.edu.pl}, \and S. J.
Lewandowski$\mathrm{^b}$ \and and \framebox{L. Leonyuk$\mathrm{^c}$} \and
{\addr $\mathrm{^a}$Donetsk Physico-Technical Institute, Ukrainian National Academy of Sciences,}\\
{\addr Luxemburg St. 72, 340114 Donetsk, Ukraine}\\
{\addr $\mathrm{^b}$Instytut Fizyki Polskiej Akademii Nauk, Al. Lotnik\'{o}w 32/46,}\\
{\addr 02-668 Warszawa, Poland}\\
{\addr $\mathrm{^c}$Moscow State University, 118899 Moscow, Russia}}
\date{}
\maketitle

\begin{abstract}
We have measured the temperature dependence of gap features revealed by Andreev
reflection ($\Delta_s$) and by tunneling ($\Delta$) in the $ab$-plane of optimal and
slightly overdoped microcrystals of $\mathrm{ (BiPb)_2Sr_2Ca_2Cu_3O_{10}}$ (Bi$2223$)
with critical temperature $T_c=110-115\:$K, and $\mathrm{ Bi_2Sr_2CaCu_2O_8}$ (Bi$2212$)
with $T_c=80-84\:$K. The tunneling conductance of Bi$2223$-Insulator-Bi$2223$ junction
shows peaks at the $2\Delta$ gap voltage, as well as dips and broad humps at other
voltages.  In Bi$2223$, similarly to the well known Bi$2212$ spectra, the energies
corresponding to $2\Delta$, to the dip, and to the hump structure are in the ratio of
$2:3:4$. This confirms that the dip and hump features are generic to the high
temperature superconductors, irrespective of the number of CuO$_2$ layers or the BiO
superstructure. On the other hand, in both compounds $\Delta(T)$ and $\Delta_s(T)$
dependences are completely different, and we conclude that the two entities have
different nature.
\end{abstract}

\newpage

\begin{doublespace}
\section{Introduction}
Along with the usual coherence gap $\Delta_s$, in the spectrum of quasiparticle
excitations in high-T$_c$ superconductors there appears a gap $\Delta_p$ (pseudogap),
which persists above the superconducting transition temperature $T_c$
\cite{Timusk,Tohoyama}. Pseudogap has the same $d$-symmetry as $\Delta_s$, but disappears
(more accurately: becomes indistinct) at some temperature $T^*>T_c$ \cite{Renner}. The
relationship between the pseudogap and superconductivity is far from clear
\cite{Timusk,Tohoyama}. One of the reasons appears to be that the most popular methods of
investigating the excitation spectrum in cuprates, like tunneling and angle-resolved
photoemission (ARPES), cannot distinguish between $\Delta_p$ and $\Delta_s$ without
recourse to various theoretical models. However, it is known that in the process of
Andreev reflection of an electron from the normal metal-superconductor (N-S) interface, a
Cooper pair is created in the superconductor \cite{Andreev}. This occurs only in the
presence of nonzero energy gap $\Delta_{s}$ in the superconductor. In other words, the
process of Andreev transformation of an electron-hole pair into a Cooper pair is possible
only for a reflection from the superconducting order parameter $\Delta_{s}$. In marked
contrast, the tunneling effect is sensitive to any singularity in the quasiparticle
excitation spectrum \cite{Wolf}. Therefore, the tunneling characteristics at $T<T_c$ in
general depend on joint contributions of the energy gap and pseudogap.

$d$-wave symmetry of the energy gap introduces some additional complications. Dominant
contribution to the junction conductivity in classical (Giaever) tunneling comes from
electrons with wave vectors forming a narrow, only a few degrees wide cone \cite{Wolf}.
Accordingly, tunnel junctions yield information on the gap anisotropy $\Delta({\mathbf
k})$, and the gap revealed in tunneling experiments can be expressed as $\Delta({\mathbf
k})=[ \Delta_s^2({\mathbf k}) + \Delta_p^2({\mathbf k})]^{1/2}$ \ \cite{Tallon}. In the
case of Andreev reflection from a clean N-S interface, the situation is different. The
incident electron is not scattered, but reflected back along the same trajectory. This is
true for any angle of incidence. It can be said that all incident electrons participate
in Andreev reflection on equal rights. Therefore, measurement of a single Andreev N-S
junction in principle is sufficient to determine the maximal value of the superconducting
gap $\Delta_{s}(\mathbf k)$.

In this paper we employ the above discussed characteristic features of
tunneling and Andreev spectroscopy to investigate the temperature dependence of
the energy gaps $\Delta$ and $\Delta_s$ in
$\mathrm{Bi_2Sr_2CaCu_2O_{8+\delta}}$ (Bi$2212$) and
(Bi,Pb)$_2$Sr$_2$Ca$_2$\-Cu$_3$O$_{10+\delta}$ [(BiPb)$2223$] cuprates. The
well studied Bi$2212$ has two CuO$_2$ layers per unit cell and strong
incommensurate modulation in the BiO layer \cite{Subramanian}, which
complicates the interpretation of tunneling and ARPES data. The substitution of
Bi by Pb in the (BiPb)$2223$ compound completely erases the superstructure in
the BiO layers.

Tunneling measurements were carried out on ``break junctions'' with the barrier surface
practically normal to the crystallographic axes in the base $ab$-plane of the material. In
the $c$ direction, the influence of BiO layer on the tunneling spectra is much more
pronounced. Andreev experiments were performed on S-N-S junctions. In both cases we
retained only the samples showing pure tunneling or Andreev characteristics. Temperature
dependencies of the energy gaps obtained in the two types of experiments were completely
different, and testify to fundamental differences between the "superconducting" gap
$\Delta_s$ and the $a,b$-axis quasiparticle gap $\Delta$.

\section{Sample preparation}
\begin{figure}
\begin{center}
\includegraphics{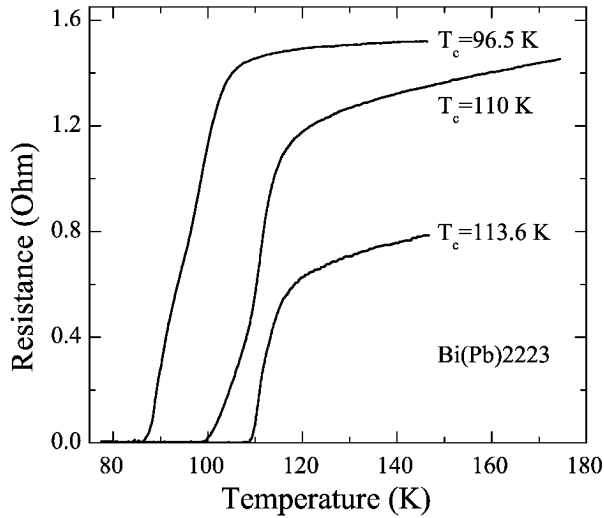}
\caption{Temperature dependence of the $ab$-plane resistance of (BiPb)$2223$ samples
with different oxygen doping.}
\end{center}
\end{figure}

The tunnel junctions were elaborated from Bi$2223$ and Bi$2212$ single crystals.
Textured $\mathrm{(Bi_{1.6}Pb_{0.4})Sr_2Ca_2Cu_2O_{10+\delta}}$ and
$\mathrm{Bi_2Sr_2CaCu_2O_8}$ samples in the form of $10\times1\times0.1 \:\mathrm{
mm^3}$ rectangular bars were prepared  \cite{JLowT,PRB,PhysC} by compacting powdered
(BiPb)$2223$ and Bi$2212$ compounds, respectively at $\mathrm{30-40\:kbar}$ between two
steel anvils. The powder was contained between two thin copper wires, whose deformation
provided uniform pressure distribution in sample volume. In this manner the powder was
compacted into dense plane-parallel bars about $0.1\,$mm thick. The bars were then
pre-annealed at $T=845\,^\circ$C for $16$ h, compressed again, and finally annealed at
$T=830\,^\circ$C for $14$~h, obtaining a well pronounced texture. Usually the Bi2212
samples were slightly overdoped and exhibited critical temperature $T_c=80-84\;$K. The
doping level of the oxygen content was controlled by annealing optimally-doped samples
in flowing gas adjusted for different partial pressures of oxygen. The samples emerging
from this procedure were highly textured, composed of tightly packed microcrystals
aligned in one direction. Sample quality was controlled by transport measurements. We
used for further processing only the samples, which were showing critical current
density $\mathrm{J_c(T=4.2\:K) > 4\cdot10^4\:A/cm^2}$. The superconducting transition
temperature $T_c$ was determined from the midpoint of the resistive $R(T)$ transition
(see Fig.~1).

The S-I-S and S-N-S junctions were made by breaking specially prepared Bi$2212$ and
Bi$2223$ samples. Each sample was hermetically sealed by insulating resin and glued to
an elastic steel plate, which was then bent until a crack occurred, running across the
sample width and detected by monitoring the sample resistance. The hermetic seal
remained unbroken in this process. After relieving the external load, the sample
returned to its initial position with the crack closed and the microcrystals once again
tightly pressed into each other on the line of the fracture. The best alignment is
expected in the sample region in which the shear deformation was minimal. This is
apparently one of the reasons, why such a procedure results in the realization of one
effective junction of the microcrystal-microcrystal type. The selection, from among the
competing junctions, of a single junction with minimal tunneling resistance is further
assisted by the nature of the tunneling effect, which decreases exponentially with the
barrier thickness. Small sample thickness ($<100\;\mu$m) and relatively large size of
the microcrystals ($>10\;\mu$m) are also important factors, enhancing junction quality.
Such break junctions on microcrystals were found to be particularly effective in the
investigation of high-T$_c$ superconductors \cite{PRB}. The typical normal-state
resistance of junctions used in the present study was between a few ohms and a few tens
of ohms, and were remarkably stable.

\begin{figure}
\begin{center}
\includegraphics{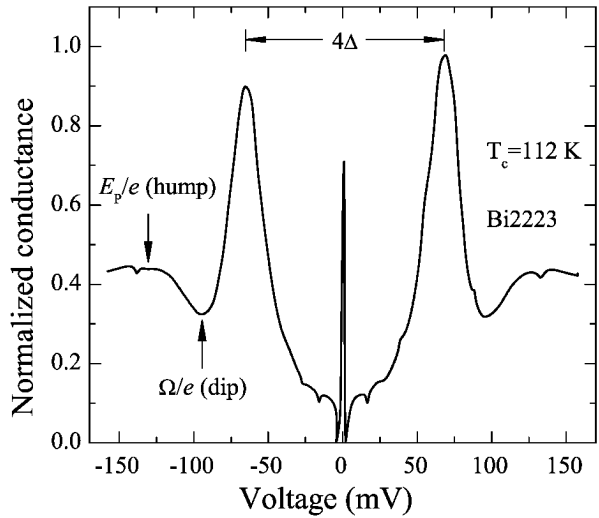}
\caption{Tunneling conductance of a Bi$2223$-I-Bi$2223$ junction at $T= 77.4\:$K. The
zero-bias peak is due to the Andreev bound state. The spectra clearly show dip and hump
structures. Arrows indicate $3\Delta$ and $4\Delta$ positions.}
\end{center}
\end{figure}

The surface of our Bi$2212$ and (BiPb)$2223$ break-junctions was perpendicular
to the CuO$_2$ plane and the direction of tunneling formed only a very small
angle $\alpha$ with one of the crystallographic axes ($a$ or $b$) in this
plane, as testified by the presence of the Andreev bound states, seen in the
tunneling S-I-S characteristics as a characteristic peak of conductivity at
zero bias (cf. Fig.~2). Numerical calculations, based on a simplified
theoretical model \cite{PRB,Tanaka} and taking into account $d$-wave mechanism
of pairing, show that the appearance of such narrow zero-bias peak in tunneling
conductance occurs at $\alpha\leq 6^\circ$. In high quality break junctions the
zero bias conductance peak (ZBCP) was reported to coexist with the Josephson
effect \cite{Cucolo}, but we have to rule out this possibility, because of
wrong signature: ZBCP was insensitive to magnetic field and did not reflect on
the I-V characteristics. The spectra $\sigma(V)=dI/dV$ show the quasiparticle
peaks at $2\Delta$, where $\Delta$ is defined as a quarter of the peak-to-peak
separation (Fig.~2). We use this $\Delta$ value as a measure of the gap, since
there is no exact method of extracting the energy gap from the tunneling
spectra, given that the exact functional form of the density of states for
high-$T_{c}$ superconductors is not known.
\begin{figure}
\begin{center}
\includegraphics{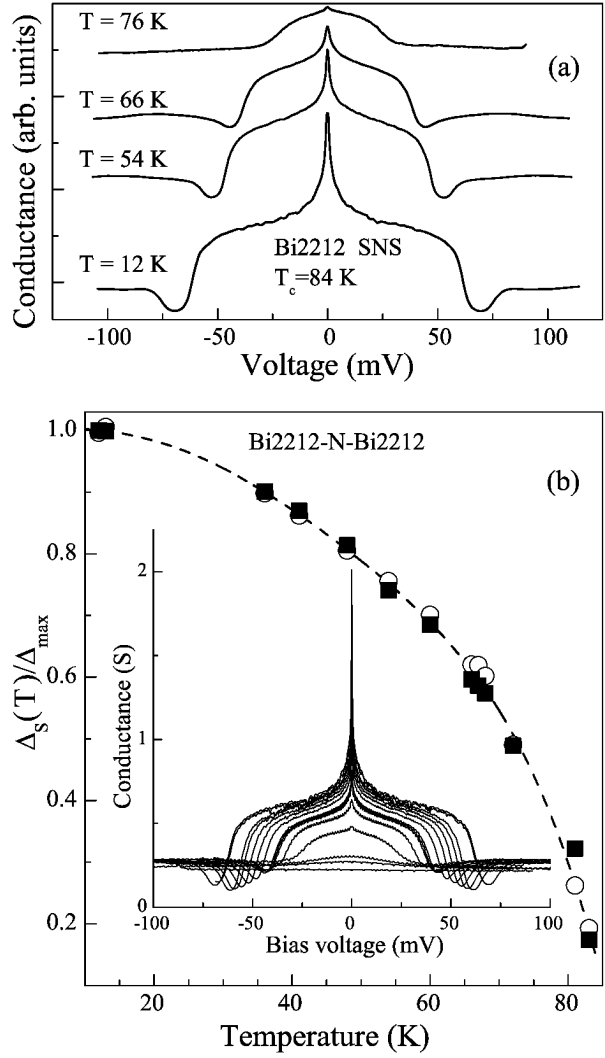}
\caption{Conductance $\sigma$ of S-N-S (Andreev) Bi$2212$-N-Bi$2212$ break junction. (a)
Temperature dependence of $\sigma$. The individual plots are shifted vertically for
clarity. (b) Temperature dependence of energy gap $\Delta_s$. The inset shows the
$\sigma$ plots in their original position.}
\end{center}
\end{figure}

In general, the type of the junction was determined \textit{ex post facto} from their
conductance $\sigma(V)$ spectra. We retained for further investigation only the
junctions conforming to either S-I-S or S-N-S types. For example, the $\sigma(V)$ curve
in Fig.~2 reveals all the characteristic features of a superconducting tunnel S-I-S
junction: an almost flat region around zero bias followed by a sharp increase in the
tunneling current, peaking around $\pm60\;$meV $(2\Delta)$; at still higher bias
voltages $V$ the conductance depends parabolically on $V$. The junction shown in Fig.~3,
on the other hand, behaves as a typical Andreev S-N-S junction. First, there is a low
resistance region at low bias voltages, seen as a broad pedestal spanning the coordinate
origin. The next indication is the excess current, which was observed in all S-N-S
junctions included in this study. Finally, the differential conductivity of the junction
at e$V>2\Delta_s$ coincides with the normal state conductivity at $T>T_c$ [see inset in
Fig.~3(b)], i.e. for $T>T_c$ practically all bias voltage is applied directly to the
junction.

One may enquire about the mechanism, which might produce in an apparently
random manner either S-I-S or S-N-S junctions. The insulating layer in S-I-S
junctions is most probably caused by oxygen depletion. As to the normal
barrier, we speculate that the CuO$_2$ planes are more hard to fracture than
the buffer layers. After breaking the sample, they
penetrate slightly into the buffer layers (see left inset in Fig.~4).
In this manner, the coupling between the CuO$_2$ planes belonging to the
separated sample parts would be stronger than the normal coupling across the
buffer layers, and it could assist to create a constriction, which would act as
a normal 3D metal. This hypothesis is in agreement with the scanning microscope
study of Bi-2212 single crystal break-junctions of the fracture surfaces, which
revealed rough, but stratified, fracture surfaces \cite{Akim}.
\begin{figure}
\begin{center}
\includegraphics{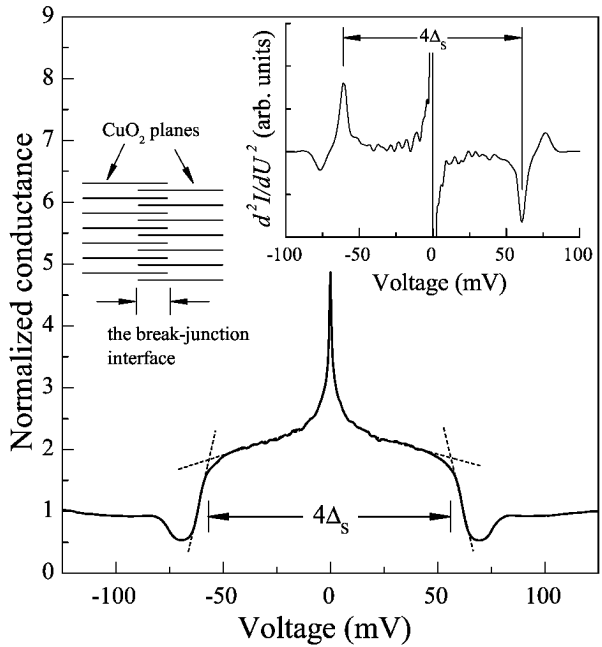}
\caption{Geometrical construction for the determination of $\Delta_s$ from Andreev
measurements. Top insert shows the corresponding $d\sigma/dV$ plot. Left insert shows
the hypothetical inner structure of the Andreev break junction.}
\end{center}
\end{figure}

\section{Experimental Results}
Temperature dependence of the energy gap $\Delta_s(T)$ obtained from Andreev S-N-S
measurements for Bi$2212$ exhibited a BCS-like form (see Fig.~3). We have used two
methods to determine $\Delta_s$ for Andreev junctions. The first one is shown in Fig.~4
and relies on measuring the distance between the points of maximal slope changes of
$\sigma(V)$ plot is taken as the measure of $4\Delta_s$. The details of the second one
are shown in top inset in Fig.~4. The rationale for both methods is in recent
calculations \cite{Dyach}, based on the Klapwijk, Blonder and Tinkham \cite{KBT}
treatment of multiple Andreev reflections between two superconductors, which indicate
that $2\Delta_s$ is determined by the separation of extrema in $d\sigma/dV$. The results
obtained by both methods are plotted together in Fig.~3(b). As seen, these results
differ slightly, but both outline essentially the same $\Delta_s(T)$ dependence.
\begin{figure}[h]
\begin{center}
\includegraphics{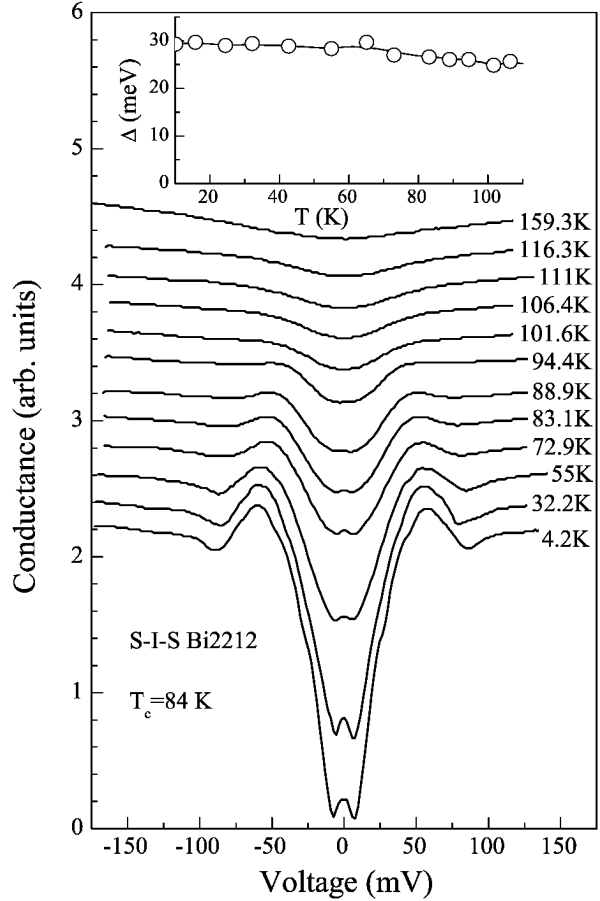}
\caption{Conductance of S-I-S (tunneling) Bi$2212$-I-Bi$2212$ break junction. Insert
shows temperature dependence of the tunneling gap $\Delta(T)$. Some structural details
of the spectra have been blurred by the speed of recording needed to overcome
temperature instabilities of the experimental setup.}
\end{center}
\end{figure}

The $\Delta(T)$ gap dependence, determined from tunneling measurements performed on the
same compounds, diverged considerably from the BCS relation (Fig.~5). In fact,
$\Delta(T)$ depends on temperature very weakly for $T\geq T_c$. According to ARPES
investigation \cite{Norman}, such behavior of $\Delta(T)$ in Bi2212 near optimal doping
is expected for the $a$ (or $b$) direction in the CuO$_2$ plane. This result agrees with
our assumption about the direction of tunneling in our Bi$2212$-I-Bi$2212$ junctions. As
mentioned above, a further confirmation is provided by the presence of Andreev bound
state, seen in the spectra of S-N-S and S-I-S junctions as a characteristic peak of
conductivity at zero bias (cf. Fig.~3 and Fig.~5). According to the ARPES data
\cite{Norman}, near optimal doping the $\Delta(T)$ gap becomes temperature dependent
only when $\alpha$ is of the order of $15^\circ$. For technological reasons, the
formation of break junctions with crystal broken at such an angle is not probable. As a
result, the tunneling characteristics at $T>T_{c}$ relate to the gap in $(100)$ or
$(010)$ direction.

In full agreement with the ARPES results \cite{Norman}, with increasing temperature the
gap $\Delta$ of Bi$2212$ becomes filled with quasiparticle excitations, and the
conductance peaks at $2\Delta$ become less distinct. The distance between the still
discernible conductance peaks does not decrease, and the $\Delta(T)$ gap is seen to
continue into the region $T>T_{c}$. Similar behavior is observed also for the
(BiPb)$2223$ compound.

The temperature dependence of the proper coherent gap $\Delta_{s}(T)$ behaves
in a completely different manner (Fig.~3). With increasing temperature, the gap
narrows, and at $T=T_c$ it closes completely. High curvature of the Andreev
conductance dip at $eV\approx 2\Delta_{s}$ is evidence both of the good quality
of the investigated junctions and of the long lifetime of quasiparticles in the
gap region. This was confirmed by the analysis of spectra of the normal metal -
constriction - superconductor (N-c-S) junctions \cite{PRB}. For Bi$2212$
Andreev N-c-S junction, the Blonder-Tinkham-Klapwijk \cite{BTK} parameter $Z$
used to obtain theoretical fit was small, $Z\simeq0.5$, a value characteristic
for very clean N-S contacts.

For energies beyond the gap $\Delta$ value, tunneling in the $ab$-plane of (BiPb)$2223$
S-I-S junction revealed the so-called dip and hump structures, as shown in Fig.~2 and
Fig.~6. In Fig.~6, the voltage axis is normalized to the voltage $eV_{p}=\Delta$, and
the conductance axis is normalized to the background; the spectra are shifted vertically
for clarity. The dip and hump features roughly scale with the gap $\Delta$ for various
oxygen doping levels (see Fig.~7). There is, however, a slight deviation of the data
from the straight $~\Delta$ line.
\begin{figure}[h]
\begin{center}
\includegraphics{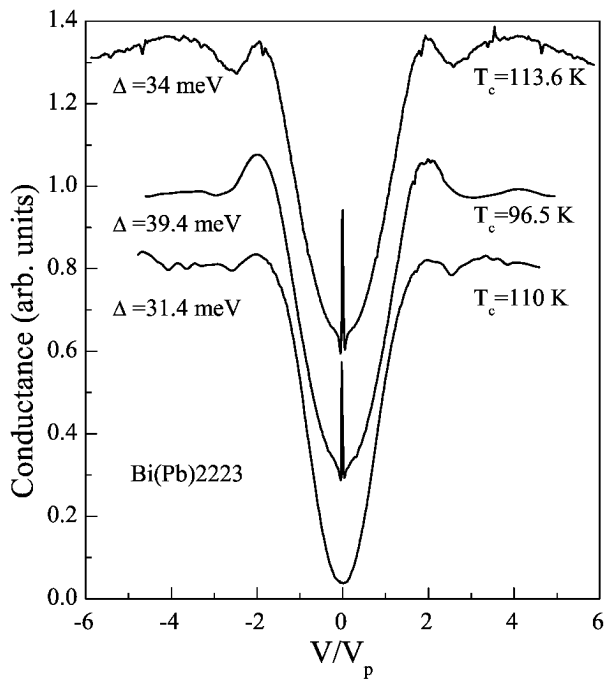}
\caption{ SIS tunneling conductance in the $ab$-plane for the Bi$2223$ samples of Fig.~1
at $T= 77.4\:$K. Voltage axis has been re-scaled in units of $\Delta$. Each curve has
been rescaled and shifted for clarity.}
\end{center}
\end{figure}

\section{Theoretical implications}
The considerable interest in the pseudogap investigation is stimulated to a
great extent by the theoretical models of high-T$_c$ superconductivity, in
which pseudogap appears as a precursor of the superconducting gap
\cite{Emery1,EKZ}, e.g. bipolaron model \cite{Alexandrov}. In another group of
models, the appearance of pseudogap is related to some sort of magnetic pairing
\cite{Anderson}. However, the domains of applicability of these models are not
very strictly defined and it is quite possible that the pseudogap (like high
temperature superconductivity) is caused by several simultaneously acting
mechanisms.
\begin{figure}
\begin{center}
\includegraphics{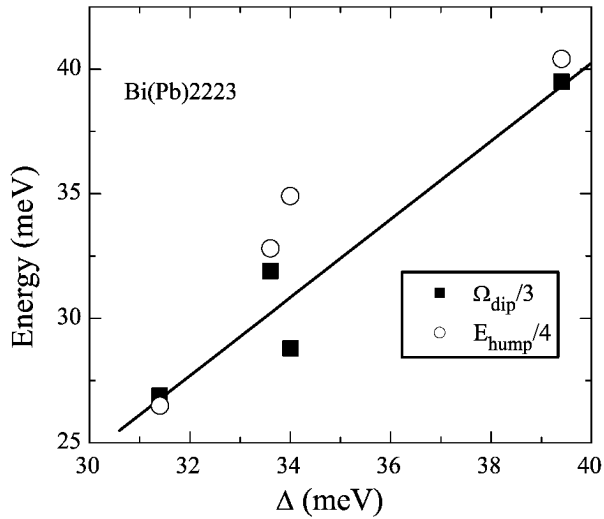}
\caption{$\Omega$ (dip) and $E_p$ (hump) positions as a function of energy gap $\Delta$
determined from the tunneling data of Fig.~2 and 6.}
\end{center}
\end{figure}

For example, in the Emerson-Kilverson-Zachar (EKZ) theory \cite{EKZ} the crucial role in
the formation of high temperature superconductivity is ascribed to the separation of
spin and charge, arising as a result of partitioning of CuO$_2$ planes into narrow
conducting and dielectric stripes. ``Pairing'' at $T^{\*}>T_c$ in the EKZ model means
the formation of a spin gap. A wide spin gap (or pseudogap $\Delta_p$) is indeed formed
in the space limited, hole-free region, such as the region between the conducting
stripes. Phase-coherent (i.e. really superconducting) state is created only at $T<T_c$.
The model well explains the smooth transition of pseudogap into tunneling gap $\Delta$
when the temperature decreases below $T_c$. However, the observed dependence on
temperature of the order parameter gap $\Delta(T)$ at $T<T_c$ is fundamentally different
from that of the gap $\Delta_s(T)$, as shown in Fig.~3 and Fig.~5. It is not clear how
in the phase fluctuation picture appears the BCS-like $\Delta_s(T)$ dependence. Such a
situation would be possible e.g. in the generation of charge (and spin) density waves
with the superconducting gap and pseudogap competing for the same region of the
Brillouin zone \cite{Markiewicz}. Then the transition to the superconducting state could
occur in the presence of a pseudogap in normal excitations, opening e.g. in the
electron-hole channel (i.e. a pseudogap, which would not transform directly into the
superconducting gap, as in the Emery-Kivelson model).

There are numerous experiments, which confirm the essentially different nature of the
superconducting gap $\Delta_{s}$ and gap (pseudogap) $\Delta$
\cite{Chen,Kabanov,Deutscher}. The most convincing are intrinsic $c$-axis tunneling
experiments (in stacked layers) \cite{Krasnov}. However, they yield different results
from the point contact, scanning tunneling spectroscopy (STM), and break junction
experiments: the hump was observed at the energy of $~2\Delta$, instead of $~4\Delta$.
The authors note a similarity between the observed $c$-axis pseudogap and Coulomb
pseudogap for tunneling into a two-dimensional electron system. In our case, the
tunneling and Andreev reflection were realized in $ab$-plane, and together with the
$\Delta(T)$ dependence (Fig.~4), we clearly observed the peak-dip-hump structure
(Figs.~2 and~3). Position of dip and hump for SIS junctions was at $~3\Delta$ and
$~4\Delta$ (Fig.~2). This suggests that the observed dip-hump structure may originate
from short-range magnetic correlations in the $ab$-plane \cite{Miyakawa}. Then the gap
$\Delta$ would be the fermionic excitation gap and $\Delta_{s}$ -- the mean-field order
parameter. It should be emphasized, finally, that the observed $\Delta_{s}(T)$
dependence exhibits non-BCS behavior at $T\rightarrow0$ (Fig.~3).

In summary, our $ab$-plane tunneling and Andreev spectroscopy studies of normal and
slightly overdoped (BiPb)$2223$ and Bi$2212$ compounds show presence both of a
superconducting energy gap $\Delta_s$, corresponding to the $d$-wave Cooper pairing, and
a dip-hump structure at $~3\Delta$ and $~4\Delta$ (for the SIS junction). This suggests
that high-energy pseudogap, which is associated with the dip and hump, could be magnetic
in origin. The gap $\Delta$ is nearly temperature independent and becomes blurred above
$T_{c}$, being continuously transformed with increasing temperature into the pseudogap.
In contrast, the order parameter gap $\Delta_{s}(T)$ has a strong temperature dependence
and for $T\rightarrow0$ reveals a non-BCS mean field behavior. Our findings are in
general agreement with those of Deutscher \cite{Deutscher}, although it must be
emphasized again that we have considered the slightly overdoped case.

\section*{Acknowledgments}
This work was supported by Polish Government (KBN) Grant No PBZ-KBN-013/T08/19.

\end{doublespace}


\begin{thebibliography}{0}
\bibitem{Timusk} T. Timusk, B. Statt, \textit{Rep. Prog. Phys.} \textbf{62}, 61 (1999).
\bibitem{Tohoyama} T. Tohoyama, S. Maekawa, \textit{Supercond. Sci. Technol.} \textbf{13}, R17 (2000).
\bibitem{Renner}  Ch. Renner, B. Revaz, J.-Y. Genoud, K. Kadowaki, {\O}. Fischer,
\textit{Phys. Rev. Lett.} \textbf{80}, 149 (1998).
\bibitem{Andreev}  A.F. Andreev, \textit{Sov. Phys. JETP} \textbf{19}, 1228 (1964).
\bibitem{Wolf}  E.L. Wolf, \textit{Principles of Electron Tunneling
Spectroscopy}, Oxford University Press, New York 1985.
\bibitem{Tallon} J.L. Tallon, G.V.M. Williams, \textit{Phys. Rev. Lett.} \textbf{82}, 3725 (1999).
\bibitem{Subramanian} M.A. Subramanian, C.C. Torardi, J.C. Calabrese, J. Gopalakrishnan,
K.J. Morrissey, T.R. Askew, R.B. Flippen, U. Chowdhry, A.W. Sleight, \textit{Science} \textbf{239}, 1015 (1988).
\bibitem{JLowT} A.I. Akimenko, T. Kita, J. Yamasaki, V.A. Gudimenko, J. Low. Temp. Phys \textbf{107}, 511 (1997).
\bibitem{PRB} A.I. D'yachenko, V.Yu. Tarenkov, R. Szymczak, A.V. Abal'oshev, I.S. Abal'osheva,
S.J. Lewandowski, L. Leonyuk, \textit{Phys. Rev. B} \textbf{61}, 1500 (2000).
\bibitem{PhysC} V.M. Svistunov, V.Yu. Tarenkov, A.I. D'yachenko, R. Aoki, \textit{Physica C} \textbf{314}, 205 (1999).
\bibitem{Cucolo} A.M. Cucolo, A.I. Akimenko, F. Bobba, F. Giubileo, Physica C \textbf{341-348}, 1589 (2000).
%\bibitem{Ekino} T. Ekino, S. Hashimoto, T. Takasai, H. Fujii, \textit{Phys. Rev
%B}, \textbf{64}, 092510 (2001)
%\bibitem{Svistuno} V.M. Svistunov, V.Yu. Tarenkov, A.I. D'yachenko, R.
%Aoki,\textit{Low Temp. Phys.}, \textbf{24}, 507 (1998).
\bibitem{Akim} A.I. Akimenko, R. Aoki, H. Murakami, V.A. Gudimenko,
\textit{Physica C} \textbf{319}, 59 (1999).
\bibitem{Dyach} A.I. D'yachenko, private communication (2002).
\bibitem{KBT} T.M. Klapwijk, G.E. Blonder, M. Tinkham, \textit{Physica B}
\textbf{109-110}, 1657 (1982).
\bibitem{Tanaka} Y. Tanaka, S. Kashiwaya, \textit{Phys. Rev. Lett.} \textbf{74}, 3451 (1995).
\bibitem{Norman} M.R. Norman, H. Ding, M. Randeria, J.C. Campuzano, T. Yokoya, T. Takeuchi,
T. Takahashi, T. Mochiku, K. Kadowaki, P. Guptasarma, D.G. Hinks , \textit{Nature} \textbf{392},
157 (1998).
\bibitem{BTK} G.E. Blonder, M. Tinkham, T.M. Klapwijk, \textit{Phys. Rev. B} \textbf{25}, 4515 (1982).
\bibitem{Emery1} V.J. Emery, S.A. Kivelson, \textit{Phys. Rev. Lett.}
\textbf{74}, 3253 (1995); \textit{Nature} \textbf{374}, 434 (1995).
\bibitem{EKZ} V.J. Emery, S.A. Kivelson, O. Zachar, \textit{Phys. Rev. B} \textbf{56}, 6120 (1997).
\bibitem{Alexandrov} A.S. Alexandrov, \textit{Philos. Trans. R. Soc. London, Ser. A}
\textbf{356}, 197 (1998), and references therein.
\bibitem{Anderson} P.W. Anderson, \textit{The Theory of Superconductivity
in the High-$T_c$ Cuprate Superconductors}, Princeton University Press, Princeton 1997.
\bibitem{Markiewicz} R.S. Markiewicz, C. Kusko,  V. Kidambi, \textit{Phys. Rev. B} \textbf{60}, 627 (1999).
\bibitem{Chen} Q. Chen, K. Levin, I. Kosztin, \textit{Phys. Rev. B} \textbf{63}, 184519 (2001).
\bibitem{Kabanov} V.V. Kabanov, J. Demsar, B. Podobnik, D.
Mihailovic, \textit{Phys. Rev. B} \textbf{59}, 1497 (1999).
\bibitem{Deutscher} G. Deutscher, \textit{Nature} \textbf{397}, 410 (1999).
\bibitem{Krasnov} V.M. Krasnov, A. Yurgens, D. Winkler, P. Delsing, T. Claeson,
\textit{Phys. Rev. Lett.} \textbf{84}, 5860 (2000).
\bibitem{Miyakawa} N. Miyakawa, J.F. Zasadzinski, L. Ozyuzer, P. Guptasarma, D.G. Hinks, C. Kendziora,
K.E. Gray, \textit{Phys. Rev. Lett.} \textbf{83}, 1018 (1999).

\end{thebibliography}
\end{document}